\documentclass[aps,prl,twocolumn,superscriptaddress]{revtex4}
\usepackage{graphicx} % include figure files 
\usepackage{bm} % bold math 
\usepackage{multirow} 
\usepackage{subfigure}
\usepackage{amsfonts}
\usepackage{color}
\begin{document}

\title{Melt-growth dynamics in CdTe crystals}

\author{X. W. Zhou}
\email[]{X. W. Zhou: xzhou@sandia.gov}
\affiliation{Sandia National Laboratories, Livermore, California 94550, USA}
%\affiliation{Mechanics of Materials Department, Sandia National Laboratories, Livermore, California 94550, USA}

\author{D. K. Ward}
\affiliation{Sandia National Laboratories, Livermore, California 94550, USA}
%\affiliation{Radiation and Nuclear Detection Materials and Analysis Department, Sandia National Laboratories, Livermore, California 94550, USA}

\author{B. M. Wong}
\affiliation{Sandia National Laboratories, Livermore, California 94550, USA}
%\affiliation{Materials Chemistry Department, Sandia National Laboratories, Livermore, California 94550, USA}

\author{F. P. Doty}
\affiliation{Sandia National Laboratories, Livermore, California 94550, USA}
%\affiliation{Radiation and Nuclear Detection Materials and Analysis Department, Sandia National Laboratories, Livermore, California 94550, USA}

\date{\today}

\begin{abstract}

We use a new, quantum-mechanics-based bond-order potential (BOP) to reveal melt-growth dynamics and fine-scale defect formation mechanisms in CdTe crystals. Previous molecular dynamics simulations of semiconductors have shown qualitatively incorrect behavior due to the lack of an interatomic potential capable of predicting both crystalline growth and property trends of many transitional structures encountered during the melt $\rightarrow$ crystal transformation. Here we demonstrate successful molecular dynamics simulations of melt-growth in CdTe using a BOP that significantly improves over other potentials on property trends of different phases. Our simulations result in a detailed understanding of defect formation during the melt-growth process. Equally important, we show that the new BOP enables defect formation mechanisms to be studied at a scale level comparable to empirical molecular dynamics simulation methods with a fidelity level approaching quantum-mechanical methods. 

\end{abstract}

% insert suggested PACS numbers in braces on next line \pacs{}

% insert suggested keywords-APS authors don't need to do this
%\keywords{}

\maketitle

% body of paper here-Use proper section commands 
% References should be done using the\cite, \ref, and \label commands

CdTe-based materials have been instrumental for technological breakthroughs in the semiconductor industry. The largest science and technology impact has been their widespread use in solar cells, radiation detectors, and medical imaging devices. While CdTe solar cells currently have the lowest cost compared to any other photovoltaic technologies \cite{Z2010}, the material is far from optimum as the record energy-conversion efficiencies achieved today are only 16\%, significantly below the 29\% theoretical value \cite{STTWK1999,DVHC2000}. CdTe-based Cd$_{1-x}$Zn$_{x}$Te (CZT) alloys are currently the leading semiconductors for $\gamma$-ray detection, but their application is limited by low manufacturing yield (and, therefore, high cost) of detector-grade materials. The under-achievement of CdTe solar cells and CZT detectors can both be attributed to charge-trapping defects formed during CdTe growth \cite{BCCCKLSJ2007,STYLBFJ2001,ZJWZZ2008}.

Direct molecular dynamics (MD) simulations provide a fundamental understanding of the CdTe growth dynamics and defect formation. However, such simulations are extremely challenging because they sample a large number of metastable configurations not known a priori. If the interatomic potential used in a simulation over-stabilizes a metastable configuration, that configuration will likely persist, leading to an amorphous growth. Previous MD simulations of semiconductor growth have shown qualitatively incorrect behavior due to the lack of an interatomic potential capable of predicting both crystalline growth and property trends of many transitional structures encountered during the growth. In addition, past MD simulations of semiconductor crystalline growth were limited to vapor deposition \cite{GFBTBR1997,XGRSBCFKSMC1994,GHR1998,SR1995,LF2000,ZMGW2006,MWZ2007}, while cases for melt-growth (which are likely to be more challenging because the crystallization occurs from a condensed liquid phase rather than the atom-by-atom assembly during vapor deposition) have yet to be demonstrated.

A vast majority of successful MD simulations of semiconductor vapor deposition \cite{GFBTBR1997,XGRSBCFKSMC1994,GHR1998,SR1995,LF2000} used Stillinger-Weber (SW) \cite{SW1985} potentials. We established previously \cite{WZWDZ2011} that while SW potentials can easily ensure the lowest energy for the equilibrium tetrahedral semiconductor crystal and its crystalline growth during vapor deposition, they cannot satisfactorily capture the property trends of other configurations. Hence, they cannot accurately reveal defect formation. Tersoff potentials \cite{T1989}, on the other hand, capture property trends more accurately. However, there is no obvious way to ensure the lowest energy for the tetrahedral structure. Because the tetrahedral structure must have a lower energy than any other structures, it is unclear which phases should be included in the potential parameterization. As a result, crystalline growth is difficult to achieve with Tersoff potentials unless the potential parameterization is done iteratively with crystalline growth used as a criterion to actively select and modify the phases to be included in the fitting. Not surprisingly, we found that many literature Tersoff potentials \cite{OG1998,NFOTMO2000,AJCHW1995} predict amorphous growth during vapor deposition simulations.

The objective of this work is to fill the missing research areas identified above by not only demonstrating the crystalline growth of semiconductor compounds from melt, but also advancing beyond Tersoff potentials on capturing properties of different phases using a new bond-order potential (BOP) \cite{PFNMZW2004,PO1999,PO2002,DMNZWP2005}. This work will begin to allow detailed investigations of defect formation mechanisms during melt- or vapor- phase growth. While we focus on CdTe compounds, the methods are applicable to a broad range of semiconductors. 

Unlike the Stillinger-Weber \cite{SW1985} and Tersoff/Brenner \cite{T1989,B1990} potentials commonly used for semiconductors, BOP is analytically derived from quantum mechanical theories \cite{PFNMZW2004,PO1999,PO2002,DMNZWP2005}. In particular, It incorporates primary ($\sigma$) and secondary ($\pi$) bonding and the valence-dependence of the heteroatom interactions, with the functional forms of the potential derived from tight-binding theory under the condition that the first two levels of the expanded Green's function are retained. Details of the parameterization of the CdTe BOP are discussed elsewhere \cite{WZWDZP2011}. This parameterization considers properties of a variety of elemental and compound configurations (with coordination from 1 to 12) including clusters, bulk lattices, defects, and surfaces, in addition to the crystalline growth of vapor deposition. 

\begin{figure}
\includegraphics[width=3in]{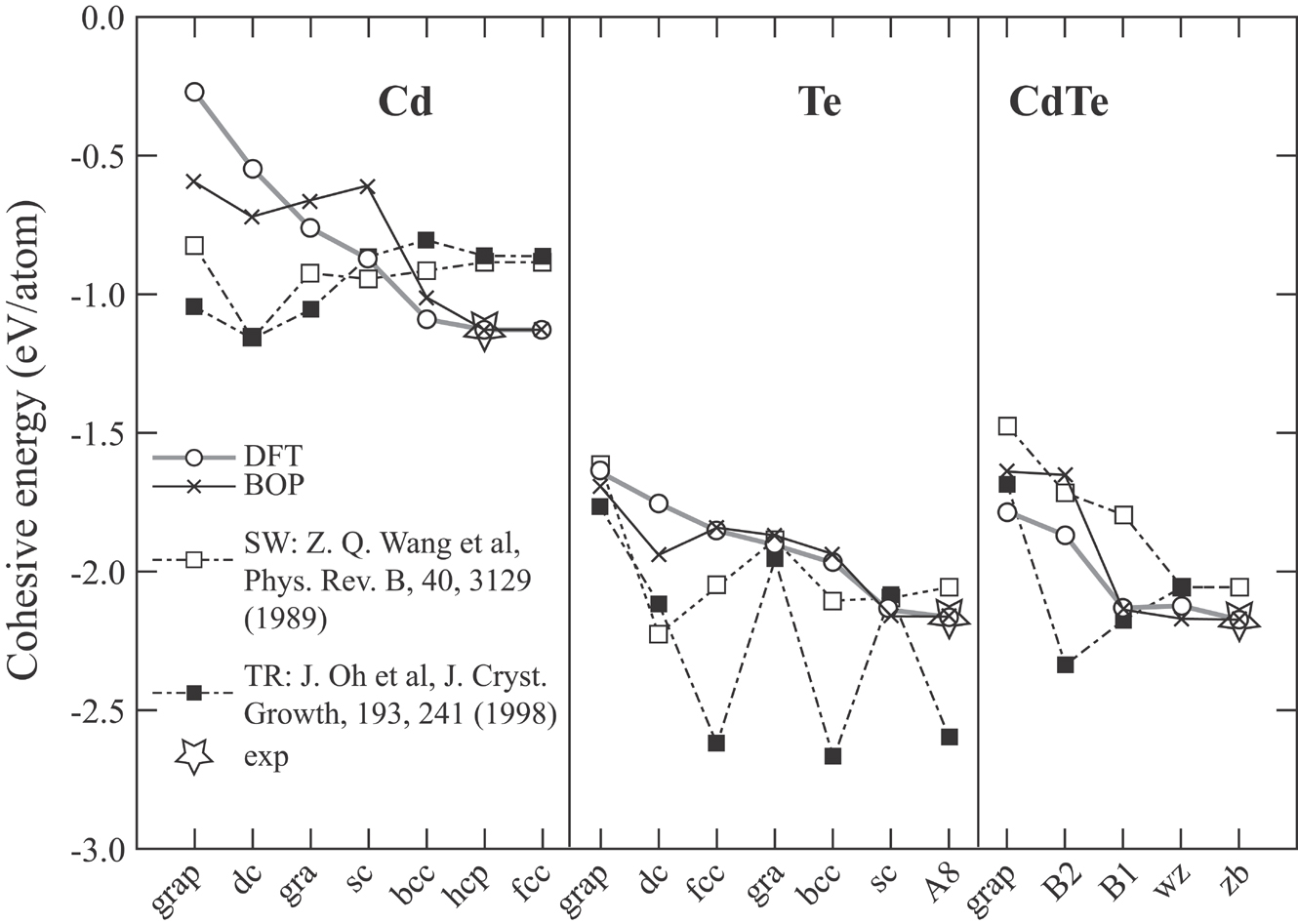}
\caption{Cohesive energies of a variety of Cd, Te, and CdTe phases calculated by various models.
\label{Ec}}
\end{figure}

There are currently two other CdTe interatomic potentials available in the literature, one \cite{WSM1989} uses the Stillinger-Weber format \cite{SW1985}, and the other \cite{OG1998} uses a Rockett modification \cite{WR1991} of the Tersoff format (TR). To evaluate different potentials, cohesive energies of a variety of Cd, Te, and CdTe phases calculated using various models are compared with the corresponding values obtained from our high-level density functional theory (DFT) calculations in Fig. \ref{Ec}. Here, various lattices are abbreviated as diamond-cubic (dc), simple-cubic (sc), body-centered-cubic (bcc), face-centered-cubic (fcc), hexagonal-close-packed (hcp), graphite (gra), graphene (grap), $\gamma$-Se (A8), zinc-blende (zb), wurtzite (wz), NaCl (B1), and CsCl (B2). For clarity, these structures are sorted to give monotonic DFT energy trends. In Fig. 1, the unfilled stars show the experimental cohesive energies \cite{B1993} of the equilibrium phases, while the straight lines connecting the neighboring data points merely guide the eye. Because the DFT calculations typically give accurate energy trends but not the absolute energies, the cohesive energies obtained from DFT calculations are scaled to match the experimental values for the equilibrium phases. Fig. 1 indicates that the cohesive energies calculated from BOP (solid lines) slightly oscillate (i.e., are non-monotonic) around the DFT benchmarks (thick light lines); however, these variations are quite minor. In fact, the BOP energy trends are considerably closer to those predicted by DFT than the corresponding results of the SW and TR parameterizations. Most importantly, BOP correctly specifies the lowest energies for the equilibrium phases of both elements and the compound, namely, the hcp Cd, the A8 Te, and the zb CdTe, and the calculated cohesive energies of the lowest energy phases also match the corresponding experimental values. In sharp contrast, the lowest energy phases are calculated to be dc Cd, dc Te, and zb CdTe by the SW parameterization and dc Cd, bcc Te, and B2 CdTe by the TR parameterization, with SW having the only correct result of zb CdTe. These results indicate that the TR parameterization cannot be used to study any of the equilibrium Cd, Te, and CdTe phases as the structures will not even be stable in MD simulations. While the SW parameterization can be used in some sort of MD simulation to study the equilibrium CdTe phase, caution should be taken in explaining the results involving defects as the potential is not transferrable to Cd and Te (and hence the defective) regimes. As a result, our new CdTe BOP approach significantly improves over other widely-used potentials on energy trends of different configurations leading to better description of defects.

\begin{figure}
\includegraphics[width=3in]{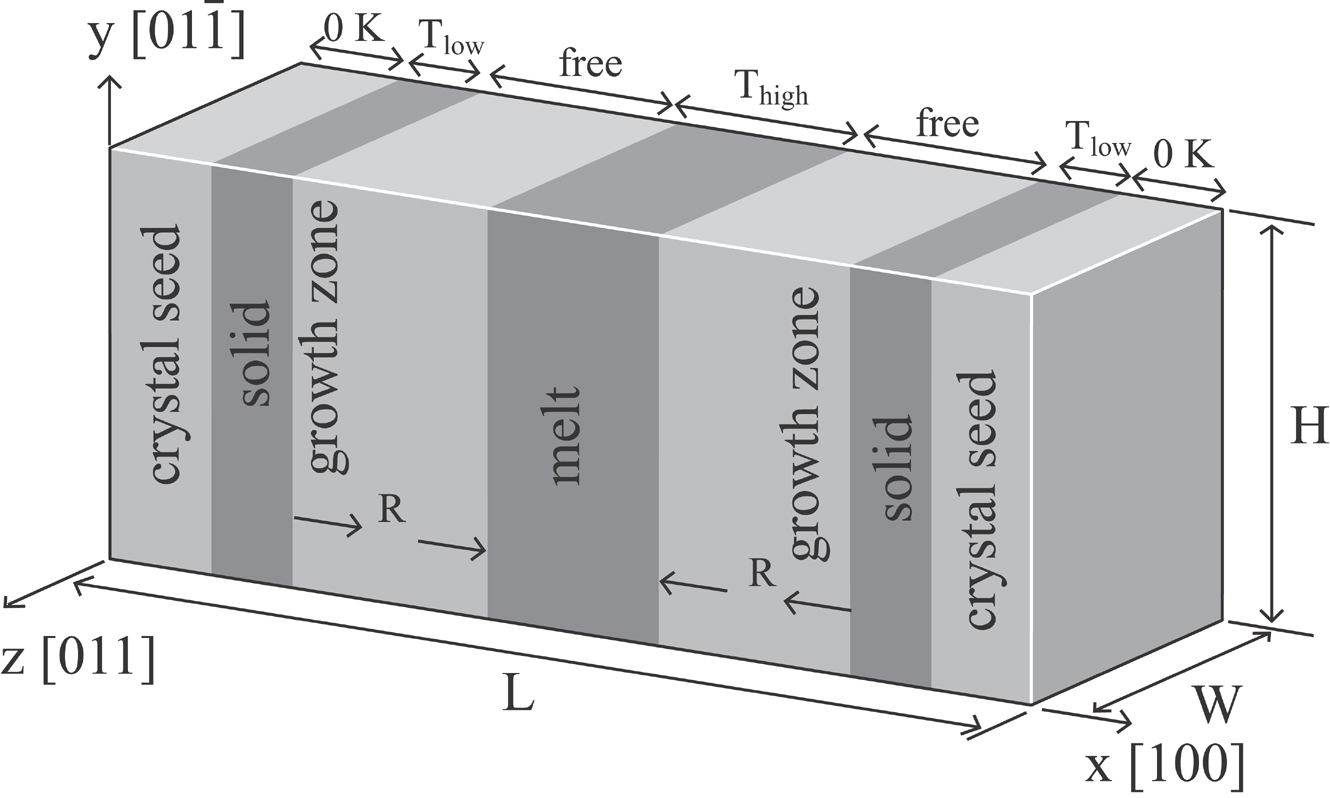}
\caption{Geometry of molecular dynamics simulations of melt-growth.
\label{model}}
\end{figure}

The geometry of the melt-growth MD simulations is shown in Fig. \ref{model}. An initial zb CdTe crystal containing 7800 Cd and 7800 Te atoms with L = 260 (400) layers (about 450 \AA) in the x- direction, H = 40 (04$\bar{4}$) layers (about 50 \AA) in the y- direction, and W = 24 (044) layers (about 30 \AA) in the z- direction was first created using the equilibrium lattice constant. Periodic boundary conditions were used in all three coordinate directions so that the system can be viewed as infinitely large. The two ends in the x- direction containing 2 (400) planes (about 1.8 \AA) were maintained at 0 K temperature (i.e., atom positions were fixed) so that these two regions acted as seeding crystals. The two regions containing 24 (400) planes (about 41 \AA) adjacent to the fixed ends were controlled at a low temperature of T$_{low}$ = 1000 K. A region containing 128 (400) planes (about 220 \AA) in the middle of the sample was controlled at a high temperature of T$_{high}$. The remainder of the sample was left free. The middle portion of the sample was then melted by first giving a random displacement to all the atoms not in the fixed regions, and then annealing the system using a MD simulation with T$_{high}$ $>$ 2200 K. The melt-growth of the crystal was then simulated in a second MD run at a different desired T$_{high}$ temperature where the T$_{low}$ and T$_{high}$ regions grew/shrank at a nominal growth rate of R = 0.2 \AA/ps, as shown in Fig. \ref{model} (the accelerated growth rate is required for MD simulations).

\begin{figure}
\includegraphics[width=3in]{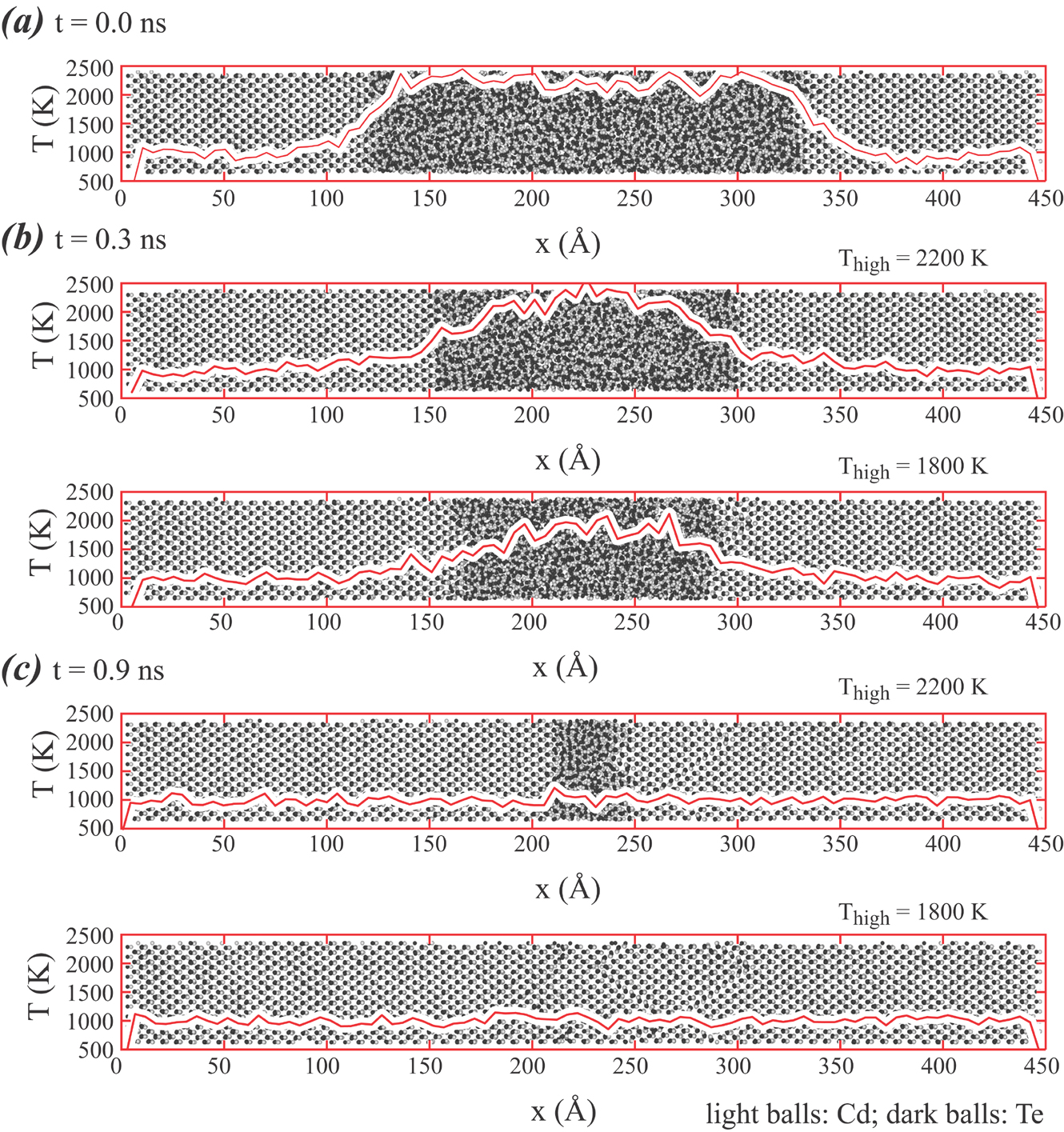}
\caption{System configurations (projected on the x-y plane) and temperature profiles (along the x- direction). (a) prior to growth; (b) 0.3 ns after growth started using two growth conditions of T$_{high}$ = 2200 K and T$_{high}$ = 1800 K; and (c) 0.9 ns after growth started using T$_{high}$ = 2200 K and T$_{high}$ = 1800 K.
\label{grow}}
\end{figure}

Simulations were performed at two different high temperatures of T$_{high}$ = 2200 K and T$_{high}$ = 1800 K. The projected x-y configurations are shown in Fig. \ref{grow} as a function of time. The temperature profiles along the x- direction are superimposed on the atomic configurations. Fig. \ref{grow}(a) shows the configuration prior to the growth where the middle section of the sample was melted at a temperature of T$_{high}$ = 2200 K. Fig. \ref{grow}(b) shows that at t = 0.3 ns, the crystal/melt interface moved to a location corresponding to a temperature T$_i$ between 1300 K and 1500 K in both the T$_{high}$ = 2200 K and T$_{high}$ = 1800 K growth conditions, in good agreement with the melting temperature of CdTe, T$_m$ = 1365 K. In contrast to conventional SW or TR-based potentials, our BOP-based MD method allows a physically-correct crystalline growth from the melt while at the same time predicts accurate energy trends of various configurations. 

Fig. \ref{grow}(c) indicates that the temperature profiles dropped to 1000 K across the sample length at t = 0.9 ns for both simulation conditions. Interestingly, the sample became entirely crystalline at T$_{high}$ = 1800 K but the middle portion remained amorphous at T$_{high}$ = 2200 K. This amorphous zone did not change with further simulation at 1000 K, but was found to crystallize when the temperature of the middle portion was raised to 1500 K and then slowly cooled to 1000 K. Fig. \ref{grow}, therefore, reveals how defects are trapped at a high growth rate.

\begin{figure}
\includegraphics[width=3in]{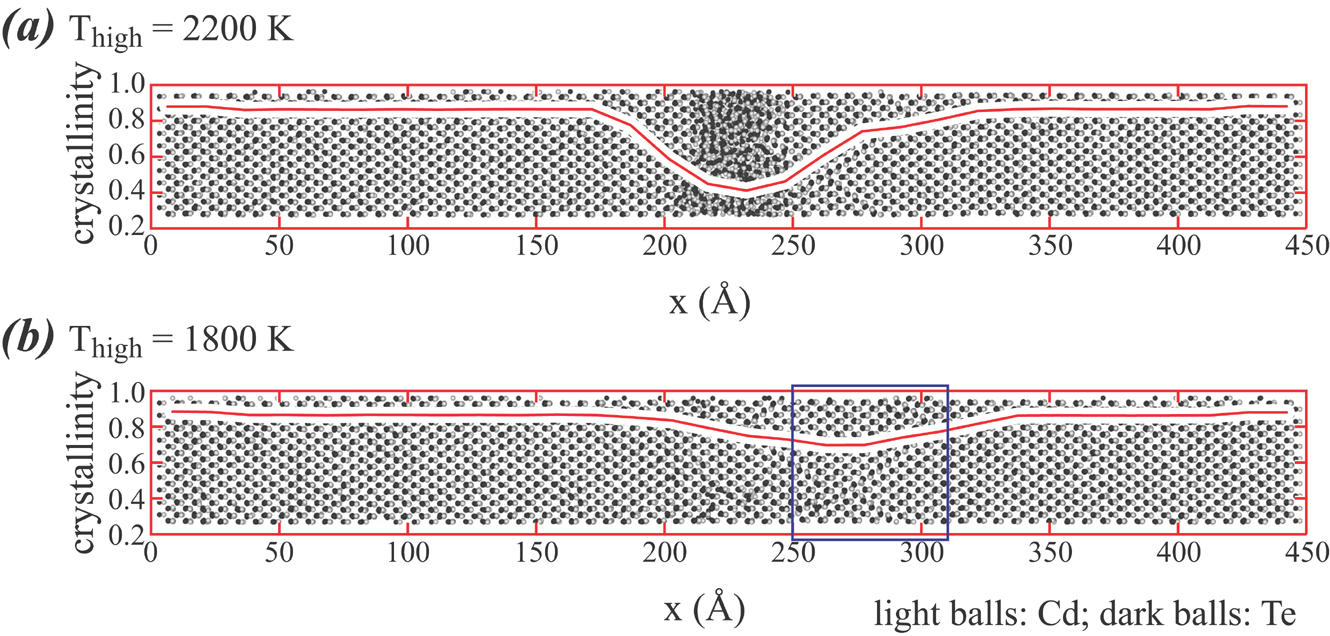}
\caption{Crystallinity analysis of the configurations shown in Fig. \ref{grow}(c). (a) T$_{high}$ = 2200 K and (b) T$_{high}$ = 1800 K. The framed region will be further analyzed in Fig. \ref{slip}.
\label{crystallinity}}
\end{figure}

\begin{figure}
\includegraphics[width=3in]{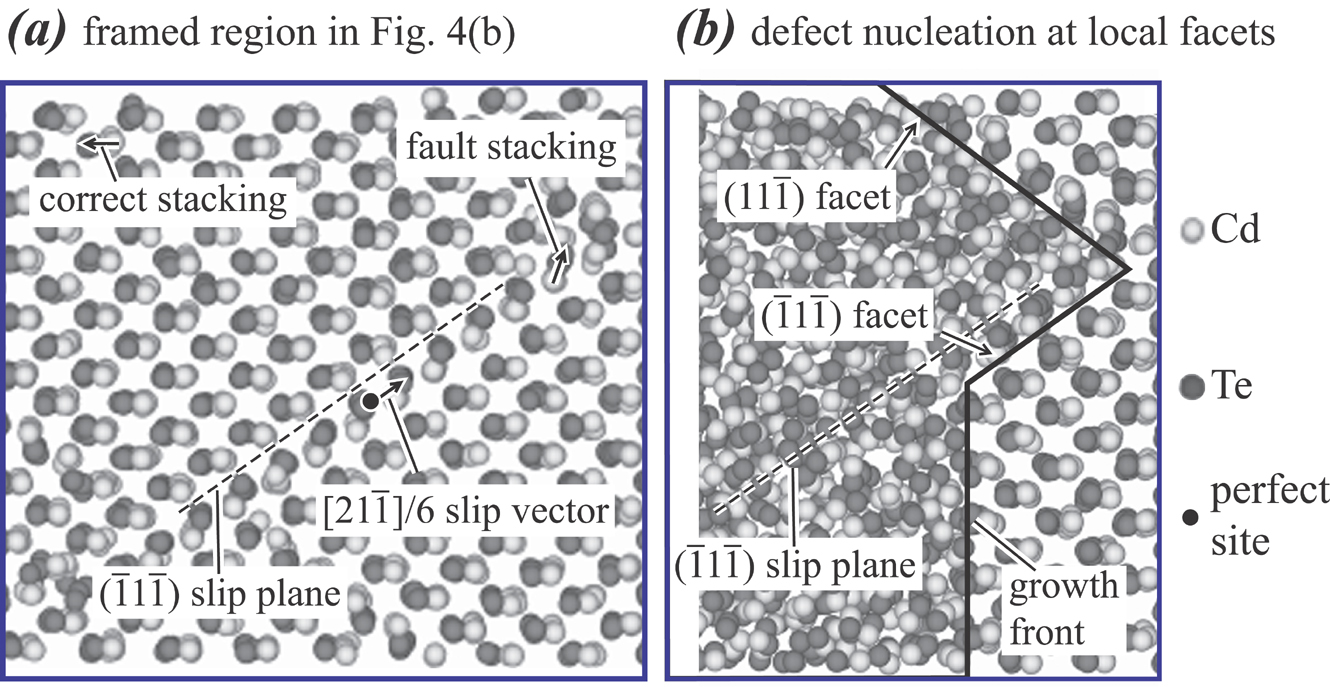}
\caption{(a) Defect analysis of the framed region shown in Fig. \ref{crystallinity}(b); and (b) Growth front at the time when the same stacking fault shown in Fig. \ref{slip}(a) was nucleated. Dashed line shows stacking fault, and thick line represents growth front.
\label{slip}}
\end{figure}

To further explore defects in the grown crystals, the crystallinity parameter developed previously \cite{ZMGW2006} was calculated along the x- direction for the two configurations shown in Fig. \ref{grow}(c), and the results are shown in Figs. \ref{crystallinity}(a) and \ref{crystallinity}(b) respectively. Fig. \ref{crystallinity}(a) shows that the crystallinity for the sample obtained from the T$_{high}$ = 2200 K condition is uniform along the sample length except between 170 \AA~and 280 \AA~where the crystallinity exhibits a drop, in agreement with the trapped amorphous phase in this region. Surprisingly, the sample obtained at T$_{high}$ = 1800 K also exhibits a crystallinity drop near the center of the sample even though no amorphous zone is identified. To understand this, the low crystallinity region, which is framed in Fig. \ref{crystallinity}(b), is magnified and examined in Fig. \ref{slip}(a). Fig. \ref{slip}(a) indicates a planar defect where the projected Cd $\rightarrow$ Te stacking, shown as an arrow in the negative x- direction, is rotated by about 109.47$^o$ clockwise. Such a rotation can originate from a [21$\bar{1}$]/6 slip on the ($\bar{1}$1$\bar{1}$) plane. The defect, therefore, corresponds to a stacking fault, and causes the drop in crystallinity observed in Fig. \ref{crystallinity}(b). Similar stacking faults have been observed in experiments \cite{SECBWZVBWN2008,HRBBW1988}. A time-resolved analysis surprisingly indicates that the growth front is not always perpendicular to the growth direction; rather, it forms local trailing \{111\} facets, suggesting a higher stability and a slower solidification on these planes. In particular, the formation of such facets was often accompanied by the nucleation of stacking faults, Fig. \ref{slip}(b). This discovery is further supported by a previous MD study on gold melt-growth, where \{111\} planes were found to cause stacking faults and slow solidification kinetics \cite{CD2002}. While mechanical twinning is known to be caused by stresses, our simulations provide a mechanistic explanation for the formation of a stacking fault: when growth occurs on a hexagonal \{111\} plane, say, plane A, atoms can occupy two different sites B and C. One of these sites corresponds to the lattice sites and the other corresponds to stacking fault sites. The energy difference between these two sites is small, and hence, it is not surprising that the condensation on a \{111\} plane may nucleate a stacking fault defect. Since the (100) and (110) planes have only one low energy (lattice) site, such stacking faults do not form if growth occurs strictly on (100) or (110) planes. These results suggest that using \{100\} or \{110\} growth planes can reduce stacking faults if the growth technique permits a sufficiently high temperature gradient and a sufficiently low growth rate to prevent the formation of local \{111\} interfaces. However, if the growth technique does not permit a sufficiently high temperature gradient and a sufficiently low growth rate, the \{111\} growth planes might be beneficial despite its high propensity for stacking faults because this interface is more likely to remain flat to minimize dendritic growth. 

In conclusion, we demonstrate that (a) new BOP-based MD simulations can accurately predict melt-growth of semiconductors; (b) strictly derived from a quantum-mechanical formalism, BOP enables defect formation mechanisms to be studied at a scale comparable to empirical MD methods and a fidelity approaching quantum-mechanical methods; and (c) amorphous defects can be trapped, and stacking faults can nucleate on \{111\} facets surprisingly formed during the melt-growth of semiconductors in non $<$111$>$ directions. 

\begin{acknowledgments}

This work is supported by the NNSA/DOE Office of Nonproliferation Research and Development, Proliferation Detection Program, Advanced Materials Portfolio. Sandia National Laboratories is a multi-program laboratory managed and operated by Sandia Corporation, a wholly owned subsidiary of Lockheed Martin Corporation, for the U.S. Department of Energy's National Nuclear Security Administration under contract DE-AC04-94AL85000.

\end{acknowledgments}

\appendix

\end{document}